# Practical Retrofitting for Obsolete Devices

Bridging the gap with old tech to create alternative interaction paradigms and workflows

Martin Lafréchoux
Montreuil, France
m.lafrechoux@gmail.com

## ABSTRACT

Over the last twenty years, smartphones gradually replaced many earlier digital tools such as PDAs, cameras and music players. Today these objects are regarded as obsolete: they may hold some esthetic or nostalgic appeal but they do not fit in a modern, zero-friction, cloud-first workflow.

Yet these devices still have desirable qualities that smartphones lack: a singular focus on a specific use case; hardware buttons and physical connectors; multi-day battery life. Even their lack of connectivity can be seen as an asset from a resilience, privacy and security standpoint.

Actually using decades-old tech today is challenging, in spite of its apparent simplicity, because the friction of physical media-based workflows now feels unacceptable. But much like classic cars can be fitted with an EV motor, it is possible to retrofit older devices in order to make them usable again in a connected world.

Long after the manufacturer stops supporting a device, user communities play a crucial role in reverse-engineering file formats and communication protocols, maintaining documentation and software archives, as well as designing and producing spare parts that can even overcome initial design flaws.

This paper will explore both software and hardware retrofitting techniques, using various examples: cameras, music players, dedicated writing instruments, video games. The resulting retrofitted devices are neither vintage nor modern, creating their own hybrid interaction paradigm around monotasking on dedicated hardware with intermittent connectivity. The various examples discussed outline some common factors that increase the likelihood that a successful retrofitting path can be found for a device. These factors can also be understood as proven design principles to create resilient hardware.

## 1 Introduction

Since the advent of the first iPhone and Android devices, the smartphone has gradually replaced many earlier analog and digital consumer devices, such as cameras and music players, much like the personal computer had replaced typewriters and various professional tools and workflows over the preceding decades. Earlier tech that fell into disuse is generally considered "obsolete". It can still hold some appeal for esthetic reasons, because it evokes a bygone, idealized era, including for people too young to have used it when it was current [26]. Yet it is not simply a matter of nostalgia: part of the appeal of vintage tech are features that actually are increasingly rare, if not unheard of, in modern devices.

Hardware:
- Physical buttons and keyboard
- Physical media support
- Standard connectors
- User-replaceable batteries
- User-serviceable design
- Sunlight readable screens, with optional lighting

Software:
- Offline-first: usable without a network connection
- Shipped complete: no software upgrades or patching necessary; no gradual software obsolescence

The smartphone is an ever-evolving glass slab that purports to replace any device. Older devices were designed once and for all to be the best tool for a specific task.

**What makes old hardware obsolete**

Old hardware can be alluring but actually using it day to day proves difficult. It is hard to fit a decades-old device in a modern workflow because communicating with it requires overcoming various challenges:

- Software issues: undocumented, proprietary file formats
- Hardware issues: repair and reliability issues, impractical physical media (e.g., floppy disks), proprietary connectors
- Connectivity issues: outdated communication protocols and hardware (dial-up modem, IrDA, unsupported Wi-Fi)

There are ways around these obstacles, as I will discuss in this paper using various examples. But the workarounds can often seem unwieldy, cumbersome, unacceptably slow. We are not used to such friction anymore: lack of connectivity feels unacceptable.[1] Constant network connectivity is the defining characteristic of modern electronics. Its absence is the fundamental reason that makes an older device "obsolete": it is impossible to patch or upgrade, cannot integrate with modern services, cannot be updated to gain new functionality. Yet network connectivity as a prerequisite is precisely what accelerates the inevitable obsolescence of newer, patchable devices.

**Paying the environmental cost**

The versatility of the smartphone comes at a high environmental price. Huge data centers are necessary to keep network-enabled tech working, yet modern devices such as smartphones actually lose functionality over time because they are *de facto* smart clients relying on an external, invisible infrastructure, leading to rapid upgrade cycles [1]. My 2012 iPad 2 is now considered vintage by Apple and cannot install apps anymore.[2] My 1992 Olivetti word processor can perform the same tasks today as when it was first released. It does not depend on an external ecosystem of services. There were no yearly software upgrades that eventually proved too much for the hardware to handle. It is as cumbersome to use as when it was first released.

Most of the environmental impact and energy use of electronic devices happens during manufacturing [25]. In a LIMITS perspective, it should be a priority to maximize the usefulness of devices that were already manufactured. Their environmental cost has already been paid in full. The most sustainable device is the one that was already manufactured decades ago and can still be made useful.

Pre-smartphone devices are better suited to this because they were designed at a time where constant network access was not a given. Tech was expensive and supposed to remain reliably useful over its lifecycle. Manufacturers now routinely ship essentially incomplete devices with the promise of future software updates to adhere to yearly upgrade cycles.[3] Older devices appear "obsolete" today precisely because they are unpatchable, but it actually makes them self-contained, resilient, and reliable. They were designed for a specific use case and thus cannot become obsolete as long as the use case exists.

Maybe the inherent friction of older devices can actually be useful, even liberating if it helps the user regain focus and purpose. How to make these devices not just usable, but actually useful today?

## 2   Background

Much of what this article discusses has already been brought up by scholars, most notably LIMITS authors.

Laurence Allard et al. discussed the environmental costs of the smartphone as a thin client relying on a huge, mostly unseen infrastructure [1]. Léa Mosesso et al. studied users of "obsolete" smartphones, their motives, the challenges they face, and the strategies they deploy in response [17].

Josh Lepawsky discussed the dwindling service life of consumer electronics, the e-waste disaster that this trend creates, as well as the role of third-party repair technicians could play in mitigating it.[10] David Franquesa et al. presented a system to reuse salvaged parts and empower users to repair their own devices [6]. Kristin N. Dew et al. studied salvage practices and proposed a series of tools enabling the reuse of salvaged materials [5].

Brian Sutherland discussed the saturation of users' drawers with discarded, obsolete devices, and degrowth strategies for computing that could break this consumerist cycle, including upcycling and considering whether a less powerful computer could not be enough [25]. Lorenz M. Hilty described the allure of the self-contained, single-purpose electronic objects of yesterday, such as the humble solar-powered pocket calculator [5]. Birgit Penzenstadler et al. showed that such autonomous, finite devices would still be useful in collapse scenarios, as opposed to the current tech paradigm [21]. Aymeric Mansoux et al. discussed the potential of minimalist devices as fertile design constraints for artists and creators in a permacomputing perspective [12]. Kelly Widdicks and Daniel Pargman described the "Cornucopian paradigm" of our current internet use, as well as strategies to moderate it [40].

What I hope to introduce is a loose practical and conceptual framework that changes the users' perspective on "obsolete" hardware and its potential usefulness, allowing users to reclaim the tools that they abandoned over the years, without them having to completely renounce modern conveniences.

## 3   Photography

At the beginning of the 21$^{st}$ century, cameras had just completed a radical transition, with digital replacing film-based products, when the modern smartphone gained prominence. In just a few years, digital cameras were absorbed by the smartphone and all but disappeared as consumer products [3]. Early digital cameras remained stuck in limbo. This multi-layered situation offers an interesting starting point for the present discussion.

### 3.1   Retro toys

The historic divide between film and digital cameras clearly marks film-based products as retro devices. There is a healthy second-hand market for film cameras as prestige products but by and large they are considered obsolete.

For the larger consumer market, several retrofit-adjacent products try to capture the retro flair of film cameras, for example modern,

---

[1] "In the same way you can never go backward to a slower computer, you can never go backward to a lessened state of connectedness" [31]
[2] As Léa Mosesso points out in her research about users with "obsolete" smartphones, this software obsolescence becomes intertwined with physical damage. The cracked screen that is considered not worth fixing becomes the physical manifestation of performances that deteriorated over time [17].
[3] This is most noticeable for smartphones, but patching culture also extends to video games or even movies that are often "fixed in post-production".

frictionless digital cameras shaped like "iconic" film devices [14]. A more surprising example is low quality plastic lenses repurposed from old disposable film cameras that can be mounted on modern digital mirrorless camera bodies through a 3D-printed housing [35].

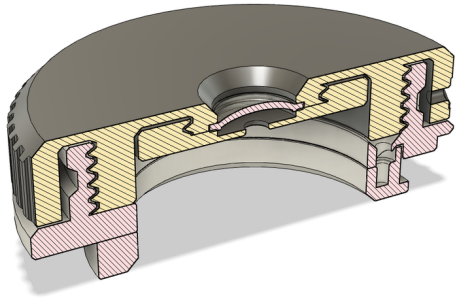

**Figure 1: A 3D-printable plastic lens housing for Sony E-Mount – © Christopher Getschmann [35]**

In both cases, these are new products that offer to bring the esthetic appeal of "retro" to a modern, network-enabled, frictionless device. While not retrofit *per se*, this shows the nostalgic value of older devices, as well as the flexibility offered by standard connectors.

## 3.2 Rediscovering earlier paradigms

Early digital cameras occupy an uneasy space between clearly obsolete products (film) and modern tech (smartphones). These cameras are functionally close enough to modern products that they do not need extensive modding to remain usable. A new battery, a flash media adapter or a new cable are often enough. But their usage has to be rediscovered.

Young Luddites love these early digital cameras precisely because they lack connectivity and any kind of app integration. To their generation this seems *unheimlich*, both strangely familiar and impossibly alien. They appreciate as novel the act of taking a picture dissociated from the urge to post it or share it [18].

## 3.3 Digital backs

When the optical part of a film camera is valuable enough, manufacturers or third parties can offer aftermarket modules that add a digital sensor to the camera body [41]. This is a straightforward retrofitting example: a new part that updates an older product designed for a different context. The original product is expensive professional equipment, has already worked for decades, and its users have no desire to move on. The resulting device is a hybrid that maintains earlier affordances while escaping the retro trap.

## 3.4 Portable scanners

In the 2024 movie *Civil War*, novice war photographer Jessie uses a portable negative scanner to digitize the film she just shot with a vintage camera and developed in the field [36]. Though the film is a fiction, the tech shown exists and is readily available to consumers. The scanner acts as a bridge between an analog, seemingly outdated workflow and a modern, network-enabled smartphone.

This choice is a leitmotiv of the movie: Jessie's travel kit grew from her personal desire to use her father's film camera. Veteran photographer Lee is initially dubious of what she perceives as an awkward gimmick but cannot help being impressed with the results.

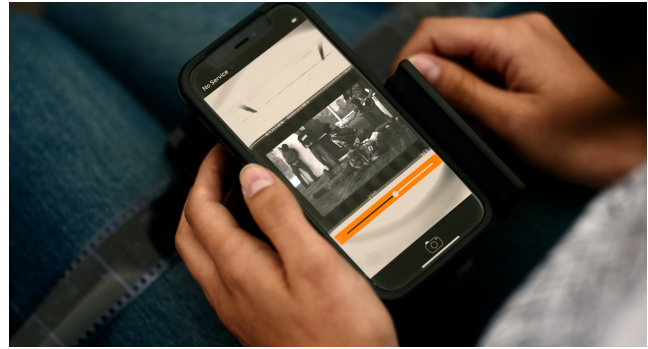

**Figure 2:** *Civil War* **(Alex Garland, 2024) – © DNA Films**

The movie makes a point of showing that digital images are plentiful and can be deleted at will. When Jessie shoots on film, friction and irreversibility give more weight and meaning to each press of the shutter.

## 4 Writing

Writing is one of the main use cases of retrofitted tech because, to paraphrase Kirby Ferguson, it is difficult to complete a novel when your typewriter is a pornography-dispensing machine [30]. Many people yearn for a device that would let them concentrate on their writing while still offering a modicum of modern convenience. Some of those people assemble in the /r/writerdeck subsection of Reddit to discuss machines whose sole purpose is to type text.

These devices include completely custom creations, new products from specialized hardware vendors, and retrofitted tech, such as word processors or educational devices.

## 4.1 AlphaSmart

AlphaSmarts were "smart keyboards" sold by AlphaSmart, Inc. from 1993 to 2005 for the educational market. They have a full-size keyboard and a small LCD screen. These devices communicate with computers by identifying as a keyboard and sending text over as bulk keypresses. A simple PS/2 to USB adapter might be needed for earlier models but they are otherwise usable today as drafting devices, at least for English speakers. International versions supporting alternate keyboard layouts and languages are difficult to source — the French model seems to have been distributed only as an accessibility device through specialized retailers, with an accordingly high price-tag. Attempts to localize the firmware have been made, to limited success.[4] The original manufacturer was acquired in 2005 and the source code never released.

## 4.2 Word processors

In the 1990s, electronic word processors from vendors such as Canon and Panasonic gained larger screens and new functionalities such as advanced layout options and spelling correction. They were effectively cheaper, basic productivity computers and came equipped with floppy drives to exchange files. These machines seem like perfect writerdecks but various issues have to be addressed.

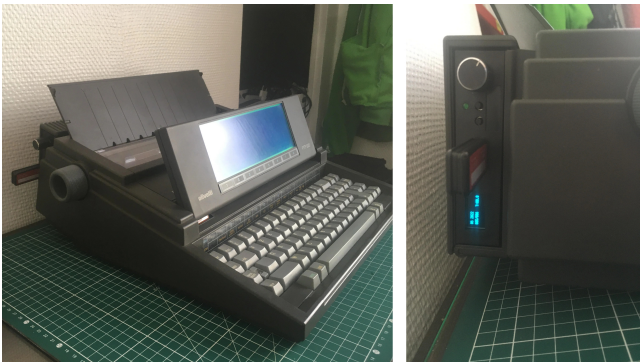

**Figure 3: A Gotek floppy emulator in an Olivetti PTP820**

*4.2.1 Floppy emulator.* Floppy disks were ubiquitous at the time but are now difficult to source and unreliable. USB floppy drives do exist but it is far more convenient to forgo the floppies entirely and replace the floppy drive with an emulator: a small, floppy-drive sized device that allows the user to plug-in a USB drive in lieu of a floppy disk. When installed, the host machine then reads and writes image files on USB drives as if they were floppy disks.[5]

*4.2.2 Encoding issues.* I bought one of these machines in 2022, an Olivetti PTP820, originally manufactured in 1992. The hardware side of the project was straightforward: the internal floppy drive used a standard connector and could be swapped for a Gotek drive. Software issues were more challenging. The Olivetti PTP-820 can save files as TXT with pure ASCII encoding or in a custom, undocumented format to accommodate French diacritics. In that case it uses a non-standard text encoding since the device predates the wide adoption of Unicode. I had to write a Python script to convert these files to UTF-8 in order to open them on my computer. Getting files from my computer back to the word processor proved more difficult: the files require a special encoding format and header that I have not yet been able to recreate fully.

*4.2.3 Wireless communication.* A further step allows to connect the word processor to a network. A tiny Linux computer such as a Raspberry Pi Zero W is configured in "USB gadget mode" to mount a file and expose it as a mass-storage device through its USB port. This computer is plugged into the Gotek USB port, exposing floppy images to the word processor, while connecting to a local Wi-Fi network in order to sync the data with a NAS or a cloud service.[6]

## 4.3 Psion 5MX

The Psion 5 is a PDA in a palmtop format from the UK firm Psion originally released in 1997. It has a delicate five-line keyboard, a wide monochrome LCD screen with a front light, and several productivity apps. An expansion port and an internal Compact Flash reader can be used to exchange files. It also has an IrDA port for wireless connectivity which is basically useless today.

More than 25 years after its initial release, the Psion 5 series still has undeniable appeal, as shown by the number of people who have worked over the years to ensure that it remains usable.[7] Multiple attempts were made to launch modernized versions (Android phones with a similar hardware keyboard in a palmtop format), even involving the original hardware designer of the Psion 5MX [20]. These products failed to reverse the trend of phones becoming black, buttonless glass slabs.

*4.3.1 Power.* The Psion has two internal power sources: a pair of AAA batteries and a backup CR2032 battery. This setup provided 15 to 20 hours of battery life [42] and allowed the user to hot swap the main batteries. Thanks to battery tech advancement, both main and backup batteries can now be replaced with Li-ion rechargeable batteries, giving the Psion far greater battery life than originally (about 30 hours of continuous use in my case, and weeks on standby).

The Psion can also be powered from an external 6V power adapter,[8] but does not have an internal charging circuit. In 2004, Matthue L. Gera documented a hardware modification to add an internal charging circuit [8]. I have not performed this modification on my device.

*4.3.2 Screen.* The Psion 5 series had a well-documented design flaw: the screen flex cable deteriorates over time with each opening and closing of the device, eventually rendering the screen unreadable. This problem was later solved by third parties: UK-based maintenance specialist Psionex designed and manufactured an upgraded, reinforced flex cable that was sold on its website [43]. I bought one in 2017 but the repair seemed too difficult to try myself. I eventually sent my Psion to a retired maintenance specialist who performed the upgrade on my device.

In December 2024, bs0dd.net published Gerber files that let anyone manufacture the reinforced flex cables [44].

---

[4] The most advanced firmware hack modifies the characters displayed on the device screen to mimic another keyboard layout, but the characters sent to the computer still use the original keyboard layout [13].
[5] Gotek floppy emulators were originally developed as a solution to retrofit industrial machines. Their stock firmware is buggy and limited but an open-source alternative firmware called FlashFloppy is available [7].

[6] This is beyond the scope of this article but replacing "the cloud" with a NAS and an open-source syncing solution such as SyncThing allows more flexibility to accommodate older, less sophisticated devices.
[7] As well as recurring articles and videos about "using one today" e.g. [15].
[8] For convenience, USB cables with a compatible barrel plug and voltage boost to 6V can be bought or created, allowing the Psion to be powered from a USB port [27].

*4.3.3 Data Storage.* The internal memory of the Psion is not persistent. It can only write to a RAM disk and its contents are reset if the batteries run out. Fortunately, the Psion can also write data to a Compact Flash card thanks to an internal port. The maximum supported size is 1GB and the Psion is notoriously picky about card models [45]. Vintage, low-capacity CF cards are now hard to source so it is easier to use a CF to microSD adapter with a 1GB microSD card, or a larger microSD with 1GB volumes. These volumes can then be read on a modern computer or smartphone to exchange files.

*4.3.4 Data syncing.* The Psion was originally designed to sync with a host computer. This creates a range of challenges today.

*4.3.4.1 Sync cable.* The Psion has an extension port that can use a syncing cable to connect to RS232. These cables are difficult to source because the connector is proprietary. A recently published solution is to reuse a PCMCIA adapter to build a custom cable since these are easier to source [38]. An RS232-to-USB adapter can then be used to connect the Psion to a modern computer.

*4.3.4.2 Sync software.* Original sync software PsiWin for Windows is still usable [28]. PsiMac and MacConnect have been unusable on a current Mac for years, but new open source syncing software is currently being developed for MacOS [39].

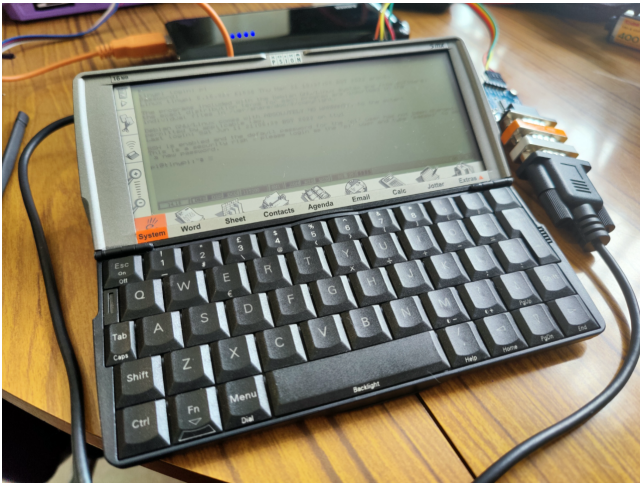

**Figure 4: a Psion 5MX using a chain of adapters to connect to the internet – © Kian Ryan [23]**

*4.3.4.3 File format.* It should also be noted that EPOC Word, the onboard word processing software, uses a proprietary file format. The easiest way to export these files in an open format is to install the utility PsiConv on the Psion itself. PsiConv was commercial software but is now considered "abandonware" by its authors and can be found on a Psion software archive [16]. This speaks to the importance of software preservation by users.

The result of all this is a pocket typewriter with a hardware keyboard, long battery life, advanced word processing software, a comfortable screen, and two-way synchronization with a host computer. The palmtop form factor has all but disappeared today and finding a Psion 5MX is often the most practical way of obtaining a device with these specifications [46].

## 4.4 MacBook Air

A November 2023 message posted on /r/writerdeck argued that the best writerdeck was simply an old laptop computer [47]. This had already been suggested many times but this user went a step further by physically removing the WiFi/Bluetooth card from his laptop to create an offline machine. I tried this with a 2011 MacBook Air from my household that needed a new battery. I was attracted to the possibility of using the note-taking application Obsidian with intermittent connectivity. As a result, the laptop now only accesses the internet when I connect my smartphone through a USB cable. Open-source utility SyncThing then syncs work and reference material folders.[9]

With this workflow, this unused laptop became my main work tool. I just plug my phone once in the morning and once at the end of the day to sync my work and research. This is the tool I used to write most of this article.

## 5 Music

Electronic music devices have long been fetishized: even though the sound and functionalities and even user interface of vintage synthesizers can now be emulated through a computer, the second-hand market suggests that many electronic musicians prefer original hardware. Famously, electronic music duo Daft Punk used exclusively vintage instruments to record their last studio album [49].

The fact that decades-old hardware is not considered obsolete, and is in fact highly sought-after, is a testament to the exceptional stability and compatibility of both the connectors and protocols used (jack and MIDI, respectively) [32]. Having no experience with electronic music creation I will focus on an example of a music-consumption and sound-recording device.

## 5.1 Sony MZ-RH1

MiniDisc (or MD) was a physical audio format created by Sony in 1992 and officially discontinued in 2012, even though Sony only stopped manufacturing blank discs in 2024. MiniDiscs store data magnetically and can be reused, like cassette tapes, yet as a digital format they offer better fidelity as well as CD-like characteristics such as track marks. MiniDisc never gained a significant market share outside of Japan because of its premium pricing but it nevertheless garnered a dedicated fan base in many other countries. In 2007, Sony released the MZ-RH1, its last ever portable MiniDisc player and recorder. It has several unique characteristics that make it an interesting retrofitting subject.

---

[9] It is worth noting that a full-fledged OS offers more flexibility to tailor the computer to one's needs and extend its useful life. Projects such as OpenCore Legacy Patcher let users install newer MacOS updates on unsupported hardware [48].

*5.1.1 Hardware.* The RH1 was designed by Sony as a swan song for the MiniDisc. It is a very compact and unapologetically premium device, with many intricate buttons, mechanical parts and trap doors that seem to take the opposite design approach of the now ubiquitous, Apple-inspired minimalism.

*5.1.1.1 Battery.* To achieve the diminutive size of the RH1, Sony used a new, very flat "gumstick" battery pack that was still user-swappable. These batteries had become difficult to source over the years, so users created 3D-printable casing that allowed alternate, standard batteries to be used as replacements [34]. Other users also designed and published 3D printed chargers and USB-C adapters for these batteries.

*5.1.1.2 Screens.* A pair of monochrome OLED screens on the front-side are one of the most striking design features of the MZ-RH1. These early OLED screens were intended as a technological showcase for Sony. Unfortunately, a not-yet-perfected manufacturing process means that these screens dim over time and eventually go dark. In 2023, Reddit user Sir68k announced he had designed a compatible replacement part and would attempt to have it manufactured. He began selling them in 2024 through online marketplace Lectronz. Each batch of a few dozens of units sells out in a few minutes. A guide and a video detailing how to install the replacement screens are provided [37]. As in the case of the Psion screen flex cable, a design flaw has been overcome by a third-party provided part. This time I did complete the procedure myself.

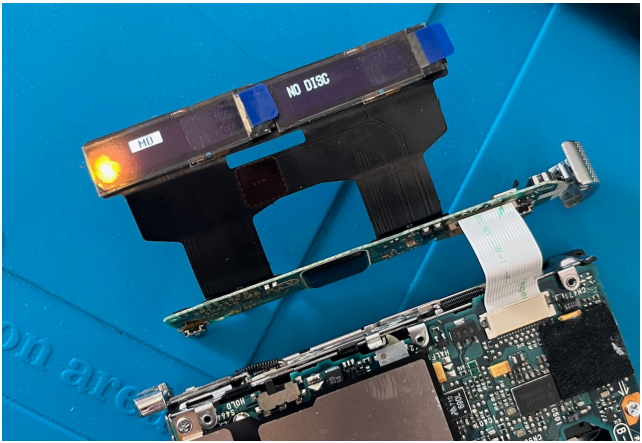

**Figure 5: installing a compatible OLED screen designed and manufactured by a third-party for the MZ-RH1**

5.1.*2* Software.

*5.1.2.1 Data import.* From December 2001, MD players branded as "NetMD" could connect to a computer through a mini USB port. This allowed users to convert sound files and burn them to a MiniDisc. The only available software to do so was a Sony-provided app, SonicStage for Windows. The sound quality of the tracks was compromised to speed up the uploading process.

When I bought an MZ-RH1 in 2013, mostly for nostalgia and esthetic value, transferring files over USB involved sourcing the last available SonicStage update from user-maintained forums and circumventing a Windows driver-signing security feature [33].

Fortunately, the communication protocol used by Sony was subsequently reverse-engineered [9] and new, open-source software was released for Windows and other platforms. Many more file types are now supported at full quality.

*5.1.2.2 Data export.* The RH1 was also the first and only MD device that allowed digital, lossless exporting of sound that had been recorded to a MiniDisc. This limitation was an artificial anti-copy measure on older MD devices and it was defeated in 2023 by the Web MiniDisc Pro developers. Now many other NetMD players can be persuaded to digitally upload the content of a disc to a computer [24].

*5.1.2.3 Firmware.* In May 2025, Sir68k announced on Reddit that he had reverse-engineered the firmware upgrade process of the MZ-RH1 and would soon release a custom firmware with new functionalities.

*5.1.3 Usage.* Thanks to the reverse-engineering efforts of a dedicated user community, the RH1 is more functional today than it was at the time of its release. Yet a lot of friction remains. MiniDiscs only hold 60 to 80 minutes of music, much like a CD, and this feels strangely limiting today. Even with new software, transferring music to a MD still takes about 25 minutes with the standard quality settings. This may seem absurd when streaming services make huge music catalogs instantly available. The album and mixtape have all but disappeared behind the infinite playlists of streaming services.

Yet taking the time to transfer CDs to MiniDisc or to create physical media from music bought on BandCamp seems to restore a form of intentionality and attention to the listening experience. Friction and limitations breed intentionality by giving weight and meaning to the user's choices.

## 5.2 Apple iPods

Similar observations can be made about Apple iPods. Hundreds of millions have been sold to consumers who generally have fond memories of the devices but do not use them anymore. Should they attempt to do so, they would quickly discover that the miniature hard drives used to store the music files and the original batteries are reaching their end-of-life, and that the syncing workflow (devised to appease the piracy fears of the music publishing industry) now feels antiquated.

The ubiquity of iPods means that compatible batteries with higher capacities are now available from third parties, as well as flash memory adapters to replace and expand failing HDDs. On the software side, a custom firmware such as Rockbox allows for a more streamlined user experience [50].

## 6 Video games

Video games offer an interesting comparison point for the present discussion because the interdependance between hardware and software creates a strong incentive for users to ensure the continuing operation of their devices.

## 6.1 Circumventing anti-piracy measures

Older consoles such as Nintendo Wii and 3DS and the Sony PSP can all be modified to run unsanctioned code. This allows the hardware to remain usable after official manufacturer support ends. This approach is most often used to run pirated games.

The commercial failure of the Sony PSP is often attributed to the early availability of custom firmwares [19]. The interests of users and creators are clearly at odds here. Software piracy is a significant economic issue that the video games industry addresses using methods sometimes so drastic that they infringe upon their customers' ability to run the games they purchased.

In the 1990s, Capcom CPS II arcade games self-destructed when their onboard battery ran out [51]. More recently, several PC game publishers used the anti-piracy tool Starforce, a piece of software so invasive that it is essentially malware [2]. In such cases the players have to devise unsanctioned methods to keep playing the games they purchased.

Video game publishers and console manufacturers appear to consider preservation efforts and long-term hardware support as a cost center offering little ROI. They prefer creating updated versions of older games that can be sold again for newer hardware. If this option does not seem worth the cost, the older game simply becomes unplayable.

## 6.2 Emulating infrastructure

The Dreamcast was the last console released by SEGA, in September 1999. Among other pioneering features it offered internet connectivity and an online service called SegaNet. Official support for this service ended in 2003 and the associated games (most notably the MMO Phantasy Star Online) became unplayable or lost functionality. Dedicated users then designed and created a replacement service called Dreamcast Live, as well as various hardware bridges to connect to it [52].

This is an interesting case that required emulating the infrastructure around the device itself to keep it working. It offers some measure of hope for the long-term life of current devices such as smartphones that were not designed to function without an extensive support infrastructure.

## 6.3 Adding new functionality

The NES (or Famicom) was the first home console released by Nintendo. It was manufactured from 1983 to 2003. The NES was designed as a modular system: throughout its lifecycle, Nintendo used various methods to expand its functionalities via special input devices, peripherals, and coprocessors integrated in the game cartridges.

In 2023, programmer Sylvain Gadrat released a new game for the NES, Super Tilt Bro, after 6 years of development and a successful crowdfunding campaign. This game supports multiplayer online play thanks to custom network code and a specially manufactured cartridge containing an ESP8266 Wi-Fi chip and an FPGA to interface with the NES hardware [53]

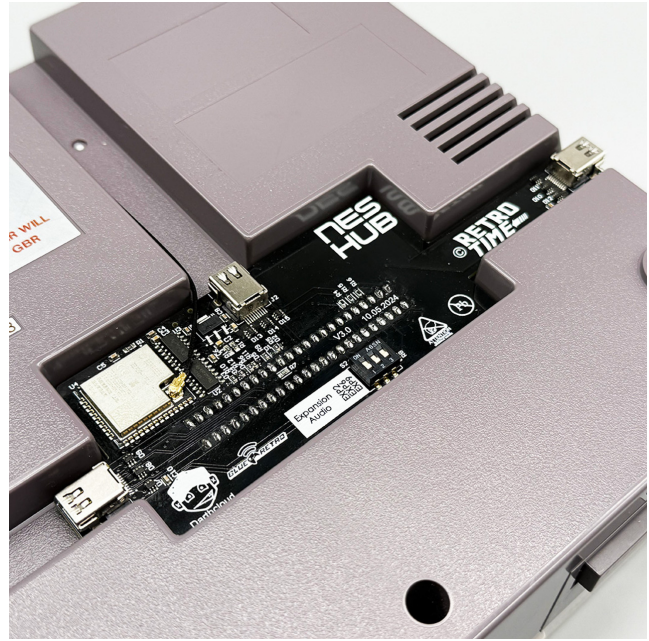

Figure 5: In 2024, modding group RetroTime released a Bluetooth adapter for the NES that plugs into an expansion port that Nintendo included but never used commercially – © 8bitmods [11]

Adding wireless network connectivity to a console 40 years after its initial release embodies the spirit of retrofitting: in bringing an old, discarded device into the present, one creates something unique and new, neither vintage nor modern, like a device out of an alternate timeline.

## 7 Discussion

The devices and use cases discussed in this paper obviously reflect my own interests and practices but the underlying processes involved can be applied to other devices and categories according to each user's needs.

**Finding a retrofitting path**

The fundamental problem to solve in order to insert an older device in a modern, network-enabled workflow is communicating with it. This does not necessarily entail a complete and direct integration with modern infrastructure and services. All that is needed is intermittent or indirect connectivity. It can be as simple as buying a USB memory card reader that supports a deprecated format. More work is usually necessary, but as long as the device has the ability to export data to a storage media, sync with a host computer or connect to a network, it should be possible to find a way to use it.

The examples discussed in this paper were chosen to highlight a wide range of problems and potential solutions:

| Problem | Potential solution |
| --- | --- |
| deprecated media | adapters, hardware emulators |
| proprietary file formats | file converters, reverse engineering |
| proprietary ports | old stock, salvaged parts, adapters |
| unavailable spare parts | old stock, salvaged parts, 3D-printed parts or adapters, third-party compatible parts |
| end of manufacturer support | user-maintained documentation and software archives |

Most, if not all of these potential solutions rely on an active, involved user community. The role of Reddit in federating such niche communities over the years cannot be overstated. Though some tinkerer and maker communities seem to increasingly favor chat apps such as Element or Discord,[10] Reddit has the advantage of being available on the open web, searchable, and extremely active (hundreds of millions of monthly users [22]), with relatively low spam content thanks to human moderation, making it the first stop for most retrofitting efforts.

**Taking inspiration from the Global South**
The second main obstacle to overcome for a successful retrofit is keeping the device in working condition over the years. Older consumer devices were not designed with yearly upgrade cycles in mind, yet they were not manufactured to be used for decades, either. Plastics decay, condensers leak, screens go dark. The internals of 1990s electronic devices are often simple enough that a moderately competent enthusiast can diagnose and fix issues using simple tools, as long as the relevant documentation is available. In many more cases, though, skilled repair technicians are paramount, as well as legislation that supports their work, as discussed by Josh Lepawsky [10].
It is also worth noting that the approach highlighted in the present article does not differ greatly from the way tech is used, maintained and repaired in developing countries. Relevant videos and tutorials to repair my 2007 laser printer came from South-East Asia, South America and African countries. In April 2025, journalists Hanan Zaffar and Danish Pandit published a feature story on tech news site The Verge exploring the laptop repair culture in Delhi, through the example of a market selling "Frankenstein laptops" assembled from second-hand computers and e-waste [29]. Salvaging parts from discarded devices to create hybrid machines that are immediately useful represents the highest possible aspiration of retrofitting.

**Designing durable hardware**
The various examples discussed in this paper also outline some common factors that increase the likelihood that a successful retrofitting path can be found.

- Open file formats

Proprietary file formats quickly become texts written in a dead language. They have to be converted to a standard, text-based format in order to be usable today.

- Standard ports

The remarkable resilience of the jack and RS232 connectors, as well as the retrocompatibility of USB through its successive iterations, are key factors in enabling communication with older devices.

- Well-established, standard communication protocols

Bluetooth has proven surprisingly useful and resilient in my retrofitting endeavors. It has been around since 1999 and BT 1.0 devices can still connect to current machines. I still use a folding keyboard from 2006 with varied hardware without issues. Wi-Fi, on the other hand, deprecates more quickly. 1990s devices such as the Apple Newton can use a PCMCIA Wi-Fi adapter, but the version of the protocol that they support is not considered secure enough by modern access points.

- Modular design

Accessible and replaceable components, especially the battery, are crucial to ensure long-term operation. Existing expansion ports can also be leveraged by third parties long after the end of manufacturer support. On the software level, an open-source firmware is obviously ideal.

- User-serviceable design

Screws and even plastic clips are much more robust and repair-friendly than adhesives.

- Long-term manufacturer support

The availability of spare parts and technical documentation is essential to long-term maintenance, and both require significant community efforts when they are lacking.

- Clarity of purpose

A device designed for a clearly defined use case cannot become obsolete as long as the use case exists.

These factors can also be understood as proven design principles to design new, resilient hardware. As an example, Nines Diamant-Berger et Barnabé Roussel, students at French engineering school UTC, are currently designing a "low-tech notebook" that shares a lot of this ethos, with the design goal of a 50-year service life [55].

---

[10] E.g., mesh networking enthusiasts rallying around Andy Kirby's Discord. [54]

It is certainly advisable that new hardware is built to last using open standards. But to ask a provocative question, do we need new hardware at all? Is there not already enough of it? Silicon manufacturing advances have been tremendous since the 1990s, but are such power and versatility necessary for the user to complete their intended task? Tech marketing always promises that newer devices will push or even abolish limits, but what if the limits are the point? This can be true for artists and creators seeking to engage with stimulating constraints, as discussed by Aymeric Mansoux et al. [12]. But any user could benefit from embracing a device that reliably does just enough to let them complete their tasks.

**Analyzing one's usage and needs**

Finding a tool that is better than the universal device requires a careful analysis of one's own needs. What exactly do I use my computer or smartphone for? Is it really the best tool for these uses, the tool that "sparks joy"?

The answer to these questions is ultimately subjective. Such interrogations often lead consumers on a quest to find the perfect device. Reddit users commonly refer to such ultimate devices as their "endgame", yet actually acquiring it rarely leads them to quietude and fulfillment. They just find a new endgame to pine for. I contend that the best tech device is the one you already have and that you only abandoned because external factors made it "obsolete". Much in the same way that people retrofit classic cars for sentimental reasons, modifying old tech to keep using it is also a way to accept, even embrace the inherent limitations of a device. There can be no disappointment or buyer's remorse.

Keeping a device for as long as possible is not some kind of sacrifice. There is nothing puritanical about it. It is actually a joyful, fulfilling process. Continually buying new gadgets in search of a perfect machine can never give the feeling of accomplishment that comes with maintaining, repairing or even enhancing the tools one already enjoys, much like a properly maintained bicycle frame that can last for a lifetime. The longer one keeps their tools, the more they become deeply personal objects whose limitations can be factored in, or even appreciated — intermittent connectivity is a blessing to the easily distracted writer, for example.

**Unbundling the smartphone**

The commonly expressed sentiment of feeling addicted to one's smartphone led to the rise of "digital detox" retreats, mindfulness apps and frameworks, both from first- and third-party vendors, and even a "dumbphone" trend [4].[11]

I argue that at least part of the problem with smartphone predominance is that it replaced other tools: it is nigh impossible to escape its pull because it is often the only tool left at our disposal. In that perspective, retrofitting old tech becomes a method for unbundling the smartphone by separating functionalities it has absorbed over the years. When one has other tools that are better suited to their purposes, the smartphone becomes less of an attention black hole and more of an access point that can allow older, offline-tech to gain an intermittent access to modern infrastructure.

## 8 Conclusion

As can be gleaned from the examples discussed, vintage tech can be modified in various ways to remain usable after it has been declared obsolete, and even gain new functionality in the process. Multiple factors drive up the likelihood that a successful retrofitting path can be found. The most obvious is the use of hardware and software standards. Proprietary ports and file formats are often the main challenges preventing the continued use of a device. A modular, expandable design is useful but not strictly necessary. Communication ports are often enough and can be utilized by sufficiently motivated users to circumvent many obstacles.

The devices discussed in this article were mostly high-end, premium products when they were first released. This has several consequences. Vintage premium devices usually enjoy a healthy second-hand market, ensuring that owners take good care of their machines. Moreover, a well-established brand selling a premium product increases the long-term availability of documentation and spare parts. If the device is still useful as a professional tool, a third-party manufacturer will provide retrofitting parts. If not, a device has to command a devoted fan base that will go to great lengths to ensure its longevity and bring it into the present.

Retrofitting old tech is neither a regression to an imagined golden era, nor blind acceptance of the consumerist promises of "progress". It is about gaining highly personal, purposeful, resilient tools that emerge organically from one's personal practice, history and needs. Users can create their own definition of progress by choosing what they want (cloud saves, better battery life, occasional network access) and not what usually comes with it (constant advertisement, attention economy, uniformization).

Retrofitted devices are hybrids that allow us to step away from the computer and smartphone and experience things on our own terms, using tools we adapted to our own purposes. They offer immediate benefits (security, sobriety) and less immediate ones (focus, purpose), but these are important too in a world where humans are expected to continually prove their value over LLMs.

There is probably no going back to a networkless age (and there never was a golden age one could go back to). But maybe we don't have to accept the modern *statu quo* wholesale. Maybe looking back at the past could allow us to try to find a way to a different future.

## ACKNOWLEDGMENTS

The author wishes to thank Aurélien Langlois and Fabienne Gallaire for their proofreading.

---

[11] The term "dumbphone" has been so widely used in recent years as to become near meaningless, now encompassing anything from a Japan-only smartphone to an old Nokia candybar phone to an Android flip phone.